\documentclass[prl,twocolumn,amsmath,amssymb]{revtex4}
\usepackage{graphicx}
\usepackage{indentfirst}
\usepackage{psfrag}
\usepackage{epsfig}
\usepackage{amsmath}
\usepackage{amssymb}
\usepackage{bm}
\usepackage{mathrsfs}
\newcommand{\be}{\begin{eqnarray}}
\newcommand{\ee}{\end{eqnarray}}

\begin{document}

\title{Comment on "Optical torque on small chiral particles in
generic optical fields " } 
\author{Manuel Nieto-Vesperinas }
\affiliation{Instituto  de Ciencia de Materiales de Madrid, Consejo Superior de
Investigaciones Cient\'{i}ficas\\
 Campus de Cantoblanco, Madrid 28049, Spain.\\ www.icmm.csic.es/mnv; 
mnieto@icmm.csic.es }



\begin{abstract}
We comment on mistakes and inaccuracies of a paper by Chen et al. concerning the optical torque from generic optical fields on dipolar chiral particles, i.e. on those whose scattering is fully described by the first electric, magnetic and magnetoelectric Mie coefficients.
\end{abstract}
\maketitle
In a recent paper \cite{chen1}  Chen et al. claim that the time-averaged optical torque exerted by a generic optical field   $ {\bf E}, {\bf B}$ on a particle, dipolar in the wide sense - i.e. such that its scattering is fully characterized by the first electric, magnetic and magnetoelectric Mie coefficients,  is given by  the sum (cf. Eq. (5)  in \cite{chen1}) of the extinction part:
\be
\frac{1}{2  } \Re [ ({\bf p}\times {\bf E}^{ *} ) + ( {\bf m}\times{\bf B}^{ *} )] ,  \label{torq2}
\ee
and the scattering  or recoil  torque, which in a Gaussian system of units reads
\be
-\frac{k^3}{3}  [\frac{1}{\epsilon}\Im( {\bf p}^{*}  \times {\bf p}) + \mu \Im ( {\bf m}^{*}  \times {\bf m}) ] \label{ttss1}.
\ee
  $\Re$ and $\Im$ stand for real and imaginary parts. ${\bf p}$ and ${\bf m}$ are the electric and magnetic dipole moments induced by the field on the particle.  $\epsilon$ and $\mu$ are the constitutive parameters of the embedding medium.

The authors of \cite{chen1} state that these above equations are taken  from   their previous work \cite{ng}, quoted as Ref. 36 in \cite{chen1}.

On the other hand, they strangely state in the introduction of \cite{chen1}  that although  the first explicit torque expression that includes the recoil terms  was reported in \cite{MNV2015_1,MNV2015_2}, "such expression is solely applicable in the case where the particle is immersed in a single plane wave field". However, they do not say that  Eqs. (\ref{torq2}) and (\ref{ttss1}) are exactly the expressions for the extinction and recoil torque that were already put forward in \cite{MNV2015_1,MNV2015_2};  i.e. before   \cite{ng} appeared. Thus,  acording to their introductory above quoted (incorrect) statement, (\ref{torq2}) and (\ref{ttss1}) would solely  apply  to plane waves.  Eqs. (6)  of \cite{chen1} are not original either, as they are already contained in \cite{,MNV2015_2}.

The problem that adds to the way these results are presented in \cite{chen1} is that while the recoil part, Eq. (\ref{ttss1}), is  valid for any arbitrary optical field, as  already stated in \cite{MNV2015_1,MNV2015_2}, the extinction component lacks a  term that should be summed to Eq. (\ref{torq2}) and that, as shown in  \cite{MNV2015_1,MNV2015_2}, is given by: 
\be
\frac{3}{4k}\sqrt{\frac{\epsilon}{\mu}}
\Im\{\frac{1}{\epsilon}({\bf p} \cdot\nabla){\bf B}^{*}-\mu  ({\bf m} \cdot\nabla){\bf E}^{ *} \}. \label{shape}
\ee
Eq. (\ref{shape}) accounts for effects of the wave  spatial  structure and polarization. This conveys both conservative and non-conservative forces associated to the electromagnetic torque \cite{MNV2015_1, MNV2015_2}.  $k= \sqrt{\epsilon\mu}\,\omega/c $.  $\omega$  being the  angular frequency, and $c$  the speed of light in vacuum. The reason  why the authors of  \cite{chen1,ng} failed to obtain Eq. (\ref{shape}) in the general expression of the torque on dipolar particles,  is due to several shortcomings in these works,  pointed out in \cite{nieto4},  which will not be repeated here.

Hence, rather than by just the sum of  (\ref{torq2}) and  (\ref{ttss1}), the correct time-averaged electromagnetic torque exerted by an arbitrary optical field on a dipolar particle should  be the sum of (\ref{torq2}),  (\ref{ttss1}) and (\ref{shape}),  which as shown in \cite{MNV2015_1, MNV2015_2} reads:
\be
<{\bm\Gamma}>=-\frac{1}{4} \Re [ ({\bf p}\times {\bf E}^{ *} ) + ( {\bf m}\times{\bf B}^{ *} )] \nonumber \\
 +\frac{3}{4k} \sqrt{\frac{\epsilon}{\mu}}\Im\{
\frac{1}{\epsilon}p_j \partial_i B_{j}^{ *}- \mu m_j \partial_i E_{j}^{ *}\}\nonumber \\
-\frac{k^3}{3}  [\frac{1}{\epsilon}\Im( {\bf p}^{*}  \times {\bf p}) + \mu \Im ( {\bf m}^{*}  \times {\bf m}) ]. \,\,\,\,\,\,\,\label{torqresuprim}
\ee
 Interestingly, addressing the time-averaged optical force $<{\bf F}>$ on a dipolar particle \cite{MNV2010}:
\begin{eqnarray}
<{\bf F}>= \frac{1}{2} \Re \{
p_j \partial_i E_{j}^{ *}+  m_j \partial_i B_{j}^{*}
-\frac{2k^4}{3}\sqrt{\frac{\mu}{\epsilon}} ({\bf p} \times {\bf m}^{*})\}, \label{force}\,\,\,\,\,\,
\end{eqnarray}
one observes reciprocal roles of  $ {\bf E}, {\bf B}$ and  ${\bf p}, {\bf m}$ in  the extinction and scattering parts of the electromagnetic torque, Eq. (\ref{torqresuprim}), with respect to those played  in the optical force, Eq. (\ref{force}).


\begin{thebibliography}{99}
\bibitem{chen1} H. Chen, W. Lu, X.  Yu, C. Xue, S. Liu, and Z. Lin, Opt. Express {\bf 25}, 32867 (2017). 
\bibitem{ng} Y. Jiang, H. Chen, J. Chen, Jack Ng and Z. Lin,  arXiv:1511.08546v2, (2016). This is the second version of  arXiv:1511.08546v1. The latter being criticised by my comment: arXiv:1605.06041.
\bibitem{MNV2015_1} M. Nieto-Vesperinas, Opt. Lett  {\bf 40}, 3021 (2015). 
\bibitem{MNV2015_2} M. Nieto-Vesperinas, Phys. Rev. A {\bf 92}, 043843 (2015).
\bibitem{nieto4} M. Nieto-Vesperinas, {\it Comment on "Universal relationships between optical force/torque and orbital versus spin momentum/angular momentum of light"},  www.researchgate.net/publication/306066037.
\bibitem{MNV2010} M. Nieto-Vesperinas, J. J. Saenz, R. Gomez-Medina, and L.
Chantada, Opt. Express {\bf  18}, 11428–11443 (2010).
\end{thebibliography}
\end{document}